\documentclass[runningheads,a4paper]{llncs}

\usepackage{amssymb}
\usepackage{placeins}

\usepackage{graphicx}

\usepackage{url}

\usepackage[utf8]{inputenc} 
\usepackage[T1]{fontenc}    
\usepackage{cite,hyperref}       
\usepackage{url}            
\usepackage{booktabs}       
\usepackage{amsfonts}       
\usepackage{nicefrac}       
\usepackage{microtype}      

\usepackage{graphicx}
\usepackage{psfrag}
\usepackage{subfigure}
\usepackage{amsmath, amssymb}
\usepackage{algorithm,algorithmic}
\usepackage{bbm}
\usepackage{etoolbox}
\usepackage{enumerate}
\usepackage{tikz}
\usetikzlibrary{shapes,arrows}
\usetikzlibrary{patterns}
\usepackage{pgfplots}
\usepackage{xspace}
\usepackage{enumerate}
\usepackage{mathtools}
\usepackage{epstopdf}
\usepackage[normalem]{ulem} 

\newtoggle{inclsupp}
\toggletrue{inclsupp}

\newtoggle{conference}
\togglefalse{conference} 

\renewcommand{\hat}{\widehat}

\def\beq{\begin{equation}}
\def\eeq{\end{equation}}
\def\beqa{\begin{eqnarray}}
\def\eeqa{\end{eqnarray}}
\def\beqan{\begin{eqnarray*}}
\def\eeqan{\end{eqnarray*}}

\def\R{{\mathbb{R}}}
\def\C{{\mathbb{C}}}

\DeclareMathOperator{\sgn}{sgn}
\DeclareMathOperator{\tr}{tr}

\DeclareMathOperator{\Cov}{Cov}

\def\xhat{\widehat{x}}

\def\zhat{\widehat{z}}

\def\Exp{\mathbb{E}}

\def\Tm1{T\! - \! 1}
\def\tm1{t\! - \! 1}
\def\tp1{t\! + \! 1}
\def\km1{k\! - \! 1}
\def\kp1{k\! + \! 1}

\newcommand{\zero}{\mathbf{0}}

\newcommand{\pbf}{\mathbf{p}}
\newcommand{\pbfhat}{\widehat{\mathbf{p}}}

\newcommand{\rbf}{\mathbf{r}}

\newcommand{\wbf}{\mathbf{w}}
\newcommand{\wbfbar}{\overline{\mathbf{w}}}

\newcommand{\xbf}{\mathbf{x}}
\newcommand{\xbfhat}{\widehat{\mathbf{x}}}
\newcommand{\xbfbar}{\overline{\mathbf{x}}}
\newcommand{\ybf}{\mathbf{y}}
\newcommand{\ybfbar}{\overline{\mathbf{y}}}
\newcommand{\zbf}{\mathbf{z}}

\newcommand{\zbfhat}{\widehat{\mathbf{z}}}
\newcommand{\Abf}{\mathbf{A}}
\newcommand{\Abfbar}{\overline{\mathbf{A}}}
\newcommand{\Bbf}{\mathbf{B}}

\newcommand{\Cbf}{\mathbf{C}}

\newcommand{\Ibf}{\mathbf{I}}

\newcommand{\thetabf}{{\boldsymbol{\theta}}}

\newcommand{\thetabfhat}{{\widehat{\boldsymbol{\theta}}}}

\newcommand{\MMSE}{_{\text{\sf MMSE}}}
\newcommand{\ML}{_{\text{\sf ML}}}

\newcommand{\tran}{^{\text{\sf T}}}

\newcommand{\defn}{\triangleq}

\newcommand*\dif{\mathop{}\!\mathrm{d}} 

\newcommand{\mat}[1]{\begin{bmatrix}#1\end{bmatrix}}

\newcommand{\pyzv}{p_{\ybf|\zbf}}

\newcommand{\pxv}{p_{\xbf}}

\newcommand{\pwv}{p_{\wbf}}

\tikzstyle{block}=[rectangle,draw, fill=blue!20,
    minimum height=1cm, minimum width=1.5cm]
\tikzstyle{signal}=[coordinate,draw]

\begin{document}

\mainmatter  

\title{An Expectation-Maximization Approach to Tuning Generalized Vector Approximate Message Passing}

\titlerunning{An Expectation Maximization Approach to Tuning GVAMP}

%
%
\author{Christopher A. Metzler \inst{1}
\and
Philip Schniter \inst{2}
\and Richard G. Baraniuk \inst{1}}
\authorrunning{An EM Approach to Tuning GVAMP}

\institute{Rice University, Department of Electrical and Computer Engineering,\\
6100 Main St. Houston, TX 77005\\ \and The Ohio State University, Department of Electrical and Computer Engineering,\\
2015 Neil Ave. Columbus, OH 43210
}
%
%

\toctitle{An EM Approach to Tuning GVAMP}
\tocauthor{An EM Approach to Tuning GVAMP}
\maketitle

\vspace{-5mm}
\begin{abstract}
Generalized Vector Approximate Message Passing (GVAMP) is an efficient iterative algorithm for approximately minimum-mean-squared-error estimation of a random vector $\mathbf{x}\sim p_{\mathbf{x}}(\mathbf{x})$ from generalized linear measurements, i.e., measurements of the form $\mathbf{y}=Q(\mathbf{z})$ where $\mathbf{z}=\mathbf{Ax}$ with known $\Abf$, and $Q(\cdot)$ is a noisy, potentially nonlinear, componentwise function. Problems of this form show up in numerous applications, including robust regression, binary classification, quantized compressive sensing, and phase retrieval. 
In some cases, the prior $p_{\mathbf{x}}$ and/or channel $Q(\cdot)$ depend on unknown deterministic parameters $\boldsymbol{\theta}$, which prevents a direct application of GVAMP.
In this paper we propose a way to combine expectation maximization (EM) with GVAMP to jointly estimate $\mathbf{x}$ and $\boldsymbol{\theta}$. 
We then demonstrate how EM-GVAMP can solve the phase retrieval problem with unknown measurement-noise variance.

\keywords{Expectation Maximization, Generalized Linear Model, Compressive Sensing, Phase Retrieval}
\end{abstract}


\vspace{-10 mm}
\section{Introduction}
\begin{figure}
\centering
\begin{tikzpicture}[scale=1]
    \node (x) {$\xbf \sim \pxv$};
    \node [block,node distance=2.0cm]  (A)   [right of=x] {$\Abf$ };
    \node [block,node distance=2.25cm]  (pyz) [right of=A] {$\pyzv$}
        edge [<-] node[auto,swap] {$\zbf$} (A);
    \node [node distance=1.65cm] (y) [right of=pyz] {$\ybf$};

    \node [below of=x,font=\footnotesize] 
        {\parbox{1.5cm}{\centering Unknown input} };
    \node [below of=A,font=\footnotesize] 
        {\parbox{1.5cm}{\centering Linear transform} };
    \node [below of=pyz,text width=1.5cm,font=\footnotesize]
        {\parbox{1.5cm}{\centering Measurement channel} };
    \node [below of=y,text width=1.5cm,font=\footnotesize,xshift=0.0cm]
        {\parbox{2.0cm}{\centering Observed measurement} };
    \draw [->] (x) -- (A);
    \draw [->] (pyz) -- (y);
\end{tikzpicture}
\vspace{-3mm}
\caption{Generalized Linear Model (GLM): 
An unknown random vector $\xbf$
is observed through a linear transform $\Abf$ 
followed by a probabilistic measurement channel $\pyzv$,
yielding the measured vector $\ybf$.  \label{fig:glm} }
\end{figure}
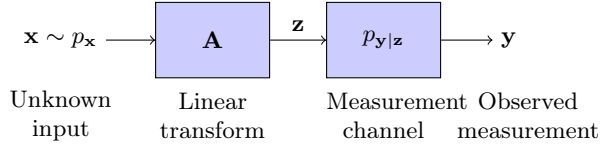

\vspace{-3mm}
We consider the problem of estimating a random vector $\xbf\in\R^N$ from observations $\ybf\in\R^M$ generated as shown in Figure~\ref{fig:glm}, which is
known as the \emph{generalized linear model} (GLM) \cite{McCulNel:89}.
Under this model, $\xbf$ has a prior density $\pxv$ and $\ybf$ obeys a likelihood function of the form $p(\ybf|\xbf)=\pyzv(\ybf|\Abf\xbf)$, where $\Abf\in\R^{M\times N}$ is a known linear transform and $\zbf\defn\Abf\xbf$ are hidden transform outputs.
The conditional density $\pyzv$ can be interpreted as a probabilistic measurement channel that accepts a vector $\zbf$ and outputs a random vector $\ybf$.
Although we have assumed real-valued quantities for the sake of simplicity, it is straightforward to generalize the methods in this paper to complex-valued quantities.

The GLM has many applications in statistics, computer science, and engineering.
For example, 
in \emph{statistical regression}, 
$\Abf$ and $\ybf$ 
contain experimental features and outcomes, respectively, and $\xbf$ are coefficients that best predict $\ybf$ from $\Abf$.
The relationship between $\ybf$ and the optimal scores $\zbf=\Abf\xbf$
is then characterized by $\pyzv$.
In \emph{imaging}-related inverse problems, 
$\xbf$ is an image to recover, $\Abf$ is often Fourier-based, and $\pyzv$ models the sensor(s).
In \emph{communications} problems, 
$\xbf$ may be a vector of discrete symbols to recover, in which case $\Abf$ is a function of the modulation/demodulation scheme and the propagation physics.
Or, $\xbf$ may contain propagation-channel parameters to recover, in which case $\Abf$ is a function of the modulation/demodulation scheme and the pilot symbols.
In both cases, $\pyzv$ models receiver hardware and interference.

Below we give some examples of the measurement channels $\pyzv$ that are encountered in these applications.
\vspace{ -2mm}
\begin{itemize}
\item \emph{Regression} 
often models $\ybf=\zbf+\wbf$ with additive noise $\wbf$, and so
$\pyzv(\ybf|\zbf)=\pwv(\ybf-\zbf)$, where $\pwv$ is the density of $\wbf$.
The ``standard linear model'' treats $\wbf$ as additive white Gaussian noise (AWGN) but is not robust to outliers.
Robust methods typically use heavy-tailed models for $\wbf$.
\item
\emph{Binary linear classification} 
can be modeled using $y_m=\sgn(z_m+w_m)$, where $\sgn(v)=1$ for $v\geq 0$ and $\sgn(v)=-1$ for $v<0$, and $w_m$ are errors.
Gaussian $w_m$ yields the ``probit'' model and logistic $w_m$ yields the ``logistic'' model.
\item
\emph{Quantized compressive sensing} 
models $y_m=Q(z_m+w_m)$, where $Q(\cdot)$ is a scalar quantizer and $w_m$ is additive, often AWGN. 
\item
\emph{Phase retrieval} 
models $y_m=|z_m|$ in the noiseless case, where $z_n\in\C$.  When noise is present, one approach is to model $y_m=|z_m+w_m|$ with $w_m\in\C$ and another is to model $y_m=|z_m|^2+w_m$ with real-valued $w_m$.
\end{itemize}
\vspace{-2mm}

In this work, we focus on the case that the prior $\pxv$ and the likelihood $\pyzv$ depend on parameters $\thetabf$ that are apriori unknown. 
For example, the prior $\pxv$ might be Bernoulli-Gaussian with unknown sparsity rate and variance, and the likelihood might involve an additive noise of an unknown variance.
We are interested in jointly estimating $\xbf$ and $\thetabf$ from $\ybf$, where $\thetabf$ are treated as deterministic.
In particular, we aim to compute the ML estimate of $\thetabf$ and the MMSE estimate of $\xbf$ under $\thetabf=\hat{\thetabf}\ML$:
\begin{subequations}\label{eq:EB}
\begin{align}
\hat{\thetabf}\ML &= \arg\max_{\thetabf} p(\ybf;\thetabf) \\
\hat{\xbf}\MMSE &= \Exp\{\xbf|\ybf;\hat{\thetabf}\ML\} 
\label{eq:MMSE},
\end{align}
\end{subequations}
sometimes referred to as the ``empirical Bayesian'' approach.

For most priors and likelihoods of interest, exact computation of the conditional mean in \eqref{eq:MMSE} is intractable.
Thus we might settle for an approximation of the MMSE estimate $\hat{\xbf}\MMSE$.
In the case that $\Abf$ is well modeled as a realization of a large rotationally invariant random matrix, the generalized vector approximate message passing (GVAMP) algorithm \cite{schniter2016vectorAsil} is a computationally efficient approach to approximate-MMSE inference under the GLM in Figure~\ref{fig:glm}.
In the large system limit (i.e., $M,N\rightarrow\infty$ with $M/N\rightarrow \delta \in (0,1)$), it is rigorously characterized by state-evolution whose fixed points, when unique, are Bayes optimal \cite{fletcher2017inference}.

For the special case of an AWGN likelihood, i.e., $\pyzv(\ybf|\zbf)=\mathcal{N}(\ybf;\zbf,\nu_w \Ibf)$ for some $\nu_w>0$,
GVAMP reduces to the simpler VAMP algorithm \cite{rangan2017vamp}.
By merging VAMP with expectation maximization (EM) \cite{DempLR:77}, one obtains the ``EM-VAMP'' approach \cite{fletcher2017learning} to the empirical-Bayesian estimation problem \eqref{eq:EB}.
In fact, with large right-rotationally invariant $\Abf$, EM-VAMP is rigorously characterized by state-evolution \cite{fletcher2017rigorous}.
Furthermore, under some identifiability conditions, it is possible to show that EM-VAMP yields an asymptotically efficient estimate of $\thetabf$.

In this paper, we propose a way to merge EM and GVAMP to tackle GLMs of the form discussed above.
This yields, for example, a way to handle phase retrieval with unknown measurement-noise variance.
The proposed ``EM-GVAMP'' approach is described in the next section. 
\vspace{-5mm}
\section{EM-GVAMP}
\vspace{-3mm}
In the sequel we assume a GLM of the form
\begin{align}
p(\ybf|\zbf;\thetabf_z) = \prod_{i=1}^M p(y_i|z_i;\thetabf_z),
\quad
\zbf = \Abf\xbf,
\quad
p(\xbf;\thetabf_x) = \prod_{j=1}^N p(x_j;\thetabf_x),
\label{eq:glm} 
\end{align}
where 
$\thetabf \defn [\thetabf_x,\thetabf_z]$ are unknown deterministic parameters,
and where $\zbf\in\R^M$ and $\xbf\in\R^N$. 

\vspace{-5mm}
\subsection{The EM Algorithm}
\vspace{-2mm}
Recalling the empirical-Bayesian methodology \eqref{eq:EB}, the maximum-likelihood estimate of $\thetabf$ given $\ybf$ can be written as
\begin{align}
\thetabfhat
&= \arg\min_\thetabf \big\{ -\ln p(\ybf;\thetabf) \big\} ,
\end{align}
where 
\begin{align}
p(\ybf;\thetabf)
&= \int p(\ybf,\zbf,\xbf;\thetabf) \dif\zbf \dif\xbf
= \int p(\ybf|\zbf,\xbf;\thetabf) p(\zbf,\xbf;\thetabf)\dif\zbf \dif\xbf\nonumber \\
&= \int p(\ybf|\zbf;\thetabf_z) \delta(\zbf-\Abf\xbf) p(\xbf;\thetabf_x) \dif\zbf \dif\xbf
\label{eq:logy} .
\end{align}
Although $p(\ybf;\thetabf)$ is difficult to work with directly, the expectation-maximization (EM) algorithm \cite{NealHinton:98} offers an alternative.
There, the idea is to write
\begin{align}
-\ln p(\ybf;\thetabf)
&= J(b;\thetabf) - D(b \,\|\, p(\xbf|\ybf;\thetabf))
\end{align}
for an arbitrary belief $b(\xbf)$, where $D(\cdot\|\cdot)$ is KL divergence, 
\begin{align}
J(b;\thetabf)
&\defn D(b \,\|\, p(\xbf;\thetabf_x)) + D(b \,\|\, p(\ybf|\xbf;\thetabf_z)) + H(b)
\end{align}
is known as the Gibbs free energy, 
and $H(b)$ is the entropy of $b$.
Because $D(b \,\|\, p(\xbf|\ybf;\thetabf))\geq 0$ for any $b$, we have that $J(b;\thetabf)$ is an upper bound on $-\ln p(\ybf;\thetabf)$, the quantity that ML seeks to minimize.
Thus, if it is tractable to construct and minimize $J(b;\thetabf)$, it makes sense to iterate the following two steps (over $k=1,2,\dots$):
\begin{align}
\text{E step:~}& b^k(\xbf)
=p(\xbf|\ybf;\thetabfhat^k)
\label{eq:Estep}\\
\text{M step:~}& \thetabfhat^{\kp1}
=\arg\min_\thetabf J(b^k;\thetabf) 
=\arg\min_\thetabf D(b^k \,\|\, p(\xbf;\thetabf_x)) + D(b^k \,\|\, p(\ybf|\xbf;\thetabf_z))
\label{eq:Mstep} ,
\end{align}
which together constitute the EM algorithm.
The ``E'' step creates an upper bound on $-\ln p(\ybf;\thetabf)$ that is tight at $\thetabf=\thetabfhat^k$, and the ``M'' step finds the estimate of $\thetabf$ that minimizes this bound.

Unfortunately, however, the posterior density required by the E-step \eqref{eq:Estep},
\begin{align}
p(\xbf|\ybf;\thetabf)
&= \frac{p(\xbf;\thetabf_x) p(\ybf|\xbf;\thetabf_z)}{p(\ybf;\thetabf)} 
= \frac{p(\xbf;\thetabf_x) \int p(\ybf|\zbf;\thetabf_z) \delta(\zbf-\Abf\xbf) \dif\zbf}{\int p(\xbf;\thetabf_x) p(\ybf|\zbf;\thetabf_z) \delta(\zbf-\Abf\xbf) \dif\zbf\dif\xbf} ,
\end{align}
is difficult to compute due to the high-dimensional integration.
Thus we consider an approximation afforded by the GVAMP algorithm \cite{schniter2016vectorAsil}.
For this, we first reparameterize the GLM \eqref{eq:glm} as a standard linear model (SLM).

\vspace{-5mm}
\subsection{An SLM Equivalent}
\vspace{-2mm}
The GLM \eqref{eq:glm} can be written as an SLM using the following formulation:
\begin{align}
\ybfbar
&= \Abfbar\xbfbar + \wbfbar
~~~\text{with}~~~
\ybfbar\defn \zero,~
\Abfbar\defn \mat{\Abf & -\Ibf_M},~
\xbfbar\defn \mat{\xbf\\\zbf},~
\wbfbar\sim \mathcal{N}(\zero,\epsilon \Ibf_M) 
~\text{s.t.}~
\epsilon\rightarrow 0 
\label{eq:slm} .
\end{align}
Here, $\xbf$ is apriori independent of $\zbf$;
the dependence between $\xbf$ and $\zbf$ manifests only aposteriori, i.e., after the measurement $\ybfbar$ is observed.
For $\xbf$, we assign the prior $p(\xbf;\thetabf_x)$, and
for $\zbf$  we assign the \emph{improper} (i.e., unnormalized) prior $p(\ybf|\zbf;\thetabf_z)$. 
The lack of normalization will not be an issue in GVAMP, because the ``prior'' $p(\ybf|\zbf;\thetabf_z)$ is used only to compute posteriors of the form 
\begin{align}
  p(\zbf|\ybf;\pbfhat,\tau,\thetabf_z) \propto 
  p(\ybf|\zbf;\thetabf_z) \mathcal{N}(\zbf;\pbfhat,\Ibf/\tau) ,
\end{align}
which are well defined because the right side is always integrable over $\zbf$.

Let us first consider direct ML estimation of $\thetabf$ in the above SLM. 
The $\thetabf$-likelihood function is 
\begin{align}
p(\ybfbar;\thetabf)
&= \int p(\ybfbar,\xbfbar;\thetabf) \dif\xbfbar
= \int p(\ybfbar|\xbfbar) p(\xbfbar;\thetabf) \dif\xbfbar
= \int \mathcal{N}(\ybfbar;\Abfbar\xbfbar,\epsilon \Ibf) p(\xbfbar;\thetabf) \dif\xbfbar \nonumber\\
&= \int \underbrace{ \mathcal{N}(\zbf;\Abf\xbf,\epsilon \Ibf) }_{\displaystyle \rightarrow \delta(\zbf-\Abf\xbf)} p(\xbf;\thetabf_x) p(\ybf|\zbf;\thetabf_z)\dif\xbf \dif\zbf 
\label{eq:logybar} ,
\end{align}
which is consistent with \eqref{eq:logy} as $\epsilon\rightarrow 0$.
Likewise, for any belief $b(\xbfbar)$, we can upper bound the negative log-likelihood by a Gibbs free energy $\bar{J}(b;\thetabf)$ of the form
\begin{align}
\bar{J}(b;\thetabf)
&\defn D(b \,\|\, p(\xbfbar;\thetabf)) + D(b \,\|\, p(\ybfbar|\xbfbar)) + H(b) 
\label{eq:Jbar} ,
\end{align}
since 
$-\ln p(\ybfbar;\thetabf)
= \bar{J}(b;\thetabf) - D(b \,\|\, p(\xbfbar|\ybfbar;\thetabf))$
with $D(b \,\|\, p(\xbfbar|\ybfbar;\thetabf))\geq 0$.
The corresponding EM algorithm is 
\begin{align}
\text{E step:~}& b^k(\xbfbar)
=p(\xbfbar|\ybfbar;\thetabfhat^k)
\label{eq:Estepbar}\\
\text{M step:~}& \thetabfhat^{\kp1}
=\arg\min_\thetabf \bar{J}(b^k;\thetabf) 
=\arg\min_\thetabf D(b^k\,\|\, p(\xbfbar;\thetabf)) 
\label{eq:Mstepbar} .
\end{align}
As before, the posterior density required by the E-step \eqref{eq:Estepbar} 
\begin{align}
p(\xbfbar|\ybfbar;\thetabf)
&= \frac{p(\xbfbar;\thetabf) p(\ybfbar|\xbfbar)}{p(\ybfbar;\thetabf)}  
= \frac{p(\xbf;\thetabf_x) p(\ybf|\zbf;\thetabf_z) \delta(\zbf-\Abf\xbf)}
       {\int p(\xbf;\thetabf_x) p(\ybf|\zbf;\thetabf_z) \delta(\zbf-\Abf\xbf) \dif\zbf\dif\xbf} ,
\end{align}
is difficult to compute due to the high-dimensional integral.
Thus we consider an approximation afforded by the GVAMP algorithm \cite{schniter2016vectorAsil}, as described in the next section.

\vspace{-5mm}
\subsection{GVAMP}
\vspace{-2mm}
Recall that the exact posterior can (in principle) be found by solving the variational optimization problem 
\begin{align}
p(\xbfbar|\ybfbar;\thetabf)
&= \arg\min_b D(b\,\|\, p(\xbfbar|\ybfbar;\thetabf)) \\
&= \arg\min_b \bar{J}(b;\thetabf) \label{eq:var1} \\
&= \arg\min_b D(b \,\|\, p(\xbfbar;\thetabf)) + D(b \,\|\, p(\ybfbar|\xbfbar)) + H(b) 
\label{eq:var2} ,
\end{align}
where \eqref{eq:var1} follows from $\bar{J}(b;\thetabf)=D(b\,\|\, p(\xbfbar|\ybfbar;\thetabf))-\ln p(\ybfbar;\thetabf)$ and \eqref{eq:var2} follows from \eqref{eq:Jbar}.
But since the posterior computation problem is NP hard in general, \eqref{eq:var2} is no more tractable than any other approach.
The GVAMP algorithm computes a posterior approximation using the expectation-consistent (EC) method \cite{OppWin:05,fletcher2016expectation}.
In this application of EC, we first split $b(\xbfbar)$ into three copies, i.e.,
\begin{align}
p(\xbfbar|\ybfbar;\thetabf)
&= \arg\min_{b_1=b_2=q} D(b_1 \,\|\, p(\xbfbar;\thetabf)) + D(b_2 \,\|\, p(\ybfbar|\xbfbar)) + H(q) ,
\end{align}
and then relax the density-matching constraint $b_1=b_2=q$ to a moment-matching constraint:
\begin{align}
p(\xbfbar|\ybfbar;\thetabf)
&\approx \arg\min_{b_1,b_2,q} D(b_1 \,\|\, p(\xbfbar;\thetabf)) + D(b_2 \,\|\, p(\ybfbar|\xbfbar)) + H(q) 
\label{eq:EC}
\\&\quad \nonumber
\text{~s.t.~} \Exp[\xbfbar|b_1]=\Exp[\xbfbar|b_2]=\Exp[\xbfbar|q] 
\text{~~and~} \tr_2\{\Cov[\xbfbar|b_1]\}\nonumber\\&\quad=\tr_2\{\Cov[\xbfbar|b_2]\}=\tr_2\{\Cov[\xbfbar|q]\} ,
\end{align}
where 
$\Exp[\xbfbar|b_i]$ and $\Cov[\xbfbar|b_i]$ denote the expectation and covariance of $\xbfbar$ under $\xbfbar\sim b_i(\xbfbar)$, and where
\begin{align}
\tr_2\left(\mat{\Abf&\Bbf\\\Bbf\tran&\Cbf}\right)
&\defn \mat{\tr(\Abf)\\\tr(\Cbf)} 
\text{~for~} \Abf\in\R^{N\times N} 
\text{~and~} \Cbf\in\R^{M\times M} .
\end{align}
Essentially, $\tr_2\{\Cov[\xbfbar]\}$ separately computes the trace of the covariance of $\xbf$ and the trace of the covariance of $\zbf$.
The right side of \eqref{eq:EC} yields three different approximations of the posterior:
\begin{align}
b_1(\xbfbar;\thetabf) 
&\propto  p(\xbfbar;\thetabf) \,\mathcal{N}\!\left(\xbfbar;\mat{\rbf_1\\\pbf_1},\mat{\Ibf_N/\gamma_1\\& \Ibf_M/\tau_1}\right) \label{eq:b1} \\
b_2(\xbfbar;\thetabf) 
&\propto  p(\ybfbar|\xbfbar) \,\mathcal{N}\!\left(\xbfbar;\mat{\rbf_2\\\pbf_2},\mat{\Ibf_N/\gamma_2\\& \Ibf_M/\tau_2}\right) \label{eq:b2} \\
q(\xbfbar;\thetabf) 
&\propto  \mathcal{N}\!\left(\xbfbar;\mat{\xbfhat\\\zbfhat},\mat{\Ibf_N/\eta\\& \Ibf_M/\zeta}\right) \label{eq:q} ,
\end{align}
where the form of \eqref{eq:b1}-\eqref{eq:q} can be deduced by analyzing the stationary points of the Lagrangian of \eqref{eq:EC}, as shown in \cite{OppWin:05}.

The GVAMP algorithm is an iterative approach to finding the values of $\rbf_1,\gamma_1,\pbf_1,\tau_1,\rbf_2,\gamma_2,\pbf_2,\gamma_2,\xbfhat,\eta,\zbfhat,\zeta$ under which the three beliefs in \eqref{eq:b1}-\eqref{eq:q} obey the moment constraints in \eqref{eq:EC}.
When $\Abf$ is large and rotationally invariant, GVAMP is rigorously characterized by a state evolution \cite{fletcher2017inference}.
Empirically, we find that the algorithm converges quickly in this scenario (e.g., on the order of $10$ iterations).

Note that the values of $\rbf_1,\gamma_1,\pbf_1,\tau_1,\rbf_2,\gamma_2,\pbf_2,\gamma_2,\xbfhat,\eta,\zbfhat,\zeta$ that satisfy the moment constraints are interdependent, and thus they all depend on the assumed value of $\thetabf$ through \eqref{eq:b1}.
\vspace{-5mm}
\subsection{EM-GVAMP}
\vspace{-3mm}
Recall that our current motivation for using GVAMP is to compute an approximation to the posterior $b^k(\xbfbar)=p(\xbfbar|\ybfbar;\thetabfhat^k)$ in the EM algorithm \eqref{eq:Estepbar}-\eqref{eq:Mstepbar}.
Of the three posterior approximations produced by GVAMP, the Gaussian approximation from \eqref{eq:q} is the simplest to use for this purpose.
Plugging the Gaussian approximation into \eqref{eq:Estepbar}-\eqref{eq:Mstepbar} yields
\begin{align}
\text{E step:~}& b^k(\xbfbar)
=\mathcal{N}\!\left(\xbfbar;\mat{\xbfhat^k\\\zbfhat^k},\mat{\Ibf_N/\eta^k\\& \Ibf_M/\zeta^k}\right) 
\text{~found via GVAMP with $\thetabf=\thetabfhat^k$}
\label{eq:approxEstepbar}\\
\text{M step:~}& \thetabfhat^{\kp1}
=\arg\min_\thetabf D(b^k\,\|\, p(\xbfbar;\thetabf)) 
\label{eq:approxMstepbar} .
\end{align}
The difference between the EM algorithm
\eqref{eq:Estepbar}-\eqref{eq:Mstepbar} 
and the EM algorithm
\eqref{eq:approxEstepbar}-\eqref{eq:approxMstepbar} 
is that, in the former case, the bound is tight at each EM iteration $k$, whereas in the latter case the bound is only approximately tight.

Due to the form of $b^k(\xbfbar)$ in \eqref{eq:approxEstepbar}, the M-step is relatively easy to compute:
\begin{align}
\thetabfhat^{\kp1}
&=\arg\min_\thetabf D(b^k\,\|\, p(\xbfbar;\thetabf)) \\
&=\arg\min_\thetabf D\big(\mathcal{N}(\xbf;\xbfhat^k,\Ibf_N/\eta^k) \mathcal{N}(\zbf;\zbfhat^k,\Ibf_M/\zeta^k)\,\big\|\, p(\xbf;\thetabf_x) p(\ybf|\zbf;\thetabf_z)\big) \\
&=\arg\max_\thetabf 
\int \mathcal{N}(\xbf;\xbfhat^k,\Ibf_N/\eta^k) \ln p(\xbf;\thetabf_x) \dif\xbf 
\nonumber\\&\quad+\int \mathcal{N}(\zbf;\zbfhat^k,\Ibf_N/\zeta^k) \ln p(\ybf|\zbf;\thetabf_z) \dif\zbf
\\
&=\arg\max_\thetabf 
\sum_{j=1}^N \int \mathcal{N}(x_j;\xhat_j^k,1/\eta^k) \ln p(x_j;\thetabf_x) \dif x_j\nonumber\\&\quad+\sum_{i=1}^M \int \mathcal{N}(z_i;\zhat_i^k,1/\zeta^k) \ln p(y_i|z_i;\thetabf_z) \dif z_i
\label{eq:approxMstepbar2} .
\end{align}
The resulting $(\thetabfhat_x^{\kp1},\thetabfhat_z^{\kp1})$ are necessarily values of $(\thetabf_x,\thetabf_z)$ that zero the gradient of the right side of \eqref{eq:approxMstepbar2} with respect to $\thetabf_x$ and to $\thetabf_z$.

\vspace{-3mm}
\section{Application to Noise-Variance Estimation in Phase Retrieval}\label{sec:EMsigmaw}
\vspace{-3mm}
In this section we will demonstrate how the EM procedure can be used to estimate noise variances in the context of phase retrieval.
Noise variance estimation in this setting has also been performed in \cite{Schniter:TSP:15} and \cite{prVBEM}. The below derivation is related to, but distinct from, these previous works.

Phase retrieval is a problem that can be formulated in the GLM setting \cite{Schniter:TSP:15}, allowing application of the GVAMP algorithm \cite{metzler2016prvamp}. We denote the special case of GVAMP applied to phase retrieval as prVAMP.

One way to model the $i$th measured intensity $y_i$ is via
\begin{align}
y_i &= \big|z_i+w_i\big|
\text{~for~i.i.d.~} w_i\sim \mathcal{N}(0,\nu_w)
\label{eq:yPR} ,
\end{align}
where $z_i,w_i\in\C$ and 
$\mathcal{N}(w_i;\mu,\nu)=\frac{1}{\pi \nu}\exp(-|w_i-\mu|^2/\nu)$
represents a circular complex-Gaussian density with mean $\mu\in\C$ and variance $\nu>0$.
In this case, the measurement noise variance $\nu_w$ may be unknown in practice, and so we might try to estimate it using the methods described in this report.
In that case, the unknown $\zbf$-likelihood parameters ``$\thetabf_z$'' reduce to $\nu_w$.
In the sequel, we will use the notation $\nu_w$ instead of $\thetabf_z$.


It was shown \cite{Schniter:TSP:15} that, under \eqref{eq:yPR}, the $z_i$-likelihood function $p(y_i|z_i;\nu_w)$ takes the form
\begin{align}
p(y_i|z_i;\nu_w)
&= 1_{y_i\geq 0}~ y_i \int_0^{2\pi} \mathcal{N}(y_i e^{j\theta_i};z_i,\nu_w) \dif\theta_i \\
&= \frac{2y_i}{\nu_w} \exp\left(-\frac{y_i^2+|z_i|^2}{\nu_w}\right) I_0\left(\frac{2y_i|z_i|}{\nu_w}\right) 1_{y_i\geq 0} ,
\end{align}
where $I_0(\cdot)$ is the $0$th-order modified Bessel function of the first kind.
If we view $p(y_i|z_i;\nu_w)$ as a density on $y_i$, then $y_i$ is Rician (conditional on $z_i$).
Note that $\theta_i$ above denotes the (hidden) phase on $z_i+w_i$; it should not be confused with the statistical parameters $\thetabf$ described earlier in this paper.

From \eqref{eq:approxMstepbar2}, we see that the EM estimate $\hat{\nu}_w^{\kp1}$ of $\nu_w$ must obey
\begin{align}
0
&=\frac{\partial}{\partial \nu_w}
\sum_{i=1}^M \int_\C \mathcal{N}(z_i;\hat{z}_i^k,1/\zeta^k) \ln p(y_i|z_i;\hat{\nu}_w^{\kp1}) \dif z_i \\
&= \sum_{i=1}^M \int_\C \mathcal{N}(z_i;\hat{z}_i^k,1/\zeta^k) 
\frac{\partial}{\partial \nu_w}
\ln \int_0^{2\pi} \mathcal{N}(y_i e^{j\theta_i};z_i,\hat{\nu}_w^{\kp1}) \dif\theta_i \dif z_i \\
&= \sum_{i=1}^M \int_\C \mathcal{N}(z_i;\hat{z}_i^k,1/\zeta^k) 
\frac{
\int_0^{2\pi} \frac{\partial}{\partial \nu_w} 
\mathcal{N}(y_i e^{j\theta_i};z_i,\hat{\nu}_w^{\kp1}) \dif\theta_i 
}{\int_0^{2\pi} \mathcal{N}(y_i e^{j\theta_i'};z_i,\hat{\nu}_w^{\kp1}) \dif\theta'_i}
\dif z_i 
\label{eq:zero} .
\end{align}
Plugging in the derivative expression (see \cite{Vila:TSP:13})
\begin{align}
\frac{\partial}{\partial \nu_w}
        \mathcal{N}(y_i e^{j\theta_i};z_i,\hat{\nu}_w^{\kp1})
&= \frac{\mathcal{N}(y_i e^{j\theta_i};z_i,\hat{\nu}_w^{\kp1})}{2(\hat{\nu}_w^{\kp1})^2}
        \big(
        |y_i e^{j\theta_i}-z_i|^2-\hat{\nu}_w^{\kp1}
        \big) 
\end{align}
into \eqref{eq:zero} and multiplying both sides by $2(\hat{\nu}_w^{\kp1})^2$, we find
\begin{align}
\hat{\nu}_w^{\kp1}
&= \frac{1}{M}\sum_{i=1}^M \int_\C \mathcal{N}(z_i;\widehat{z}_i^k,1/\zeta^k)
        \frac{ 
        \int_0^{2\pi} 
        |y_i e^{j\theta_i}-z_i|^2
        \mathcal{N}(y_i e^{j\theta_i};z_i,\hat{\nu}_w^{\kp1}) 
        \dif\theta_i}
        {\int_0^{2\pi} 
        \mathcal{N}(y_i e^{j\theta_i'};z_i,\hat{\nu}_w^{\kp1}) 
        \dif\theta_i'}
        \dif z_i \\
&= \frac{1}{M}\sum_{i=1}^M 
        \int_\C
        \mathcal{N}(z_i;\hat{z}_i^k,1/\zeta^k) 
        \int_0^{2\pi}
        |y_i e^{j\theta_i}-z_i|^2
        p(\theta_i;z_i,\hat{\nu}_w^{\kp1})
        \dif\theta_i \dif z_i  
                \label{eq:EM1}
\end{align}
with the newly defined pdf
\begin{align}
p(\theta_i;z_i,\hat{\nu}_w^{\kp1})
&\triangleq \frac{ \mathcal{N}(y_i e^{j\theta_i};z_i,\hat{\nu}_w^{\kp1}) }
        {\int_0^{2\pi} \mathcal{N}(y_i e^{j\theta_i'};z_i,\hat{\nu}_w^{\kp1}) 
        \dif\theta_i'} 
\propto 
 \exp\Big(-\frac{|z_i-y_i e^{j\theta_i}|^2}{\hat{\nu}_w^{\kp1}}\Big)\\
&\propto 
 \exp\big(\kappa_i \cos(\theta_i-\phi_i)\big)
 \text{~for~}\kappa_i \triangleq \frac{2|z_i|y_i}{\hat{\nu}_w^{\kp1}},
        \label{eq:ptheta}
\end{align}
where $\phi_i$ denotes the phase of $z_i$. 
The expression \eqref{eq:ptheta} identifies this pdf as a von Mises distribution \cite{Mardia:Book:00}, which can be stated in normalized form as
\begin{align}
p(\theta_i;z_i,\widehat{\nu}_w^{\kp1})
&= \frac{\exp(\kappa_i \cos(\theta_i-\phi_i))}{2\pi I_0(\kappa_i)} .
        \label{eq:ptheta2}
\end{align}
Expanding the quadratic in \eqref{eq:EM1} and plugging in \eqref{eq:ptheta2}, we get
\begin{align}
\widehat{\nu}_w^{\kp1}
&= \frac{1}{M}\sum_{i=1}^M 
        \int_{\C}
        \mathcal{N}(z_i;\widehat{z}_i^k,1/\zeta^k) 
        \bigg( y_i^2 + |z_i|^2 
        \nonumber\\\quad&
        - 2y_i |z_i| 
        \int_0^{2\pi}
        \cos(\theta_i-\phi_i)\,
        \frac{\exp(\kappa_i \cos(\theta_i-\phi_i))}{2\pi I_0(\kappa_i)} 
        \dif\theta_i 
        \bigg)
        \dif z_i \\  
&= \frac{1}{M}\sum_{i=1}^M 
        \int_{\C}
        \mathcal{N}(z_i;\widehat{z}_i^k,1/\zeta^k) 
        \bigg( y_i^2 + |z_i|^2  - 2y_i |z_i| 
        R_0\bigg(\frac{2|z_i|y_i}{\widehat{\nu}_w^{\kp1}}\bigg) 
        \bigg)
        \dif z_i 
\label{eq:EM2} ,
\end{align}
where $R_0(\cdot)$ is the modified Bessel function ratio 
$R_0(\kappa_i)\triangleq I_1(\kappa_i)/I_0(\kappa_i)$
and \eqref{eq:EM2} follows from \cite[9.6.19]{Abramowitz:Book:64}.

Simplifying approximations of \eqref{eq:EM2} could be taken as needed.
For example, in the high-SNR case, the expansion
$R_0(\kappa) = 1 - \frac{1}{2\kappa} - \frac{1}{8\kappa^2} - \frac{1}{8\kappa^3}+ o(\kappa^{-3})$
from \cite[Lemma~5]{Robert:SaPL:90}
could be used to justify 
\begin{align}
R_0(\kappa) 
&\approx 1 - \frac{1}{2\kappa} , 
\end{align}
which, when applied to \eqref{eq:EM2}, yields
\begin{align}
\widehat{\nu}_w^{\kp1}
&\approx \frac{2}{M}\sum_{i=1}^M 
        \int_{\mathbb{C}}
        \big( y_i - |z_i| \big)^2 
        \mathcal{N}(z_i;\widehat{z}_i^k,1/\zeta^k) 
        \dif z_i .
        \label{eq:EM3} 
\end{align}
Approximation \eqref{eq:EM3} can be reduced to an expression that involves the mean of a Rician distribution. 
In particular, using $z_i=\rho_i e^{j\phi_i}$, the integral in \eqref{eq:EM3} can be converted to polar coordinates as follows:
\begin{align}
\int_0^\infty \big( y_i - \rho_i \big)^2 
        \hspace{-11mm}
        \underbrace{
        \int_0^{2\pi} \mathcal{N}(\rho_i e^{j\phi_i};\widehat{z}_i^k,1/\zeta^k) 
        \dif \phi_i \rho_i 
        }_{\displaystyle 
\frac{2\rho_i}{1/\zeta^k} 
\exp\left(-\frac{\rho_i^2+|\hat{z}_i^k|^2}{1/\zeta^k}\right)
I_0\left(\frac{2\rho_i|\hat{z}_i^k|}{1/\zeta^k}\right)
1_{\rho_i\geq 0}
        }
        \hspace{-11mm}
\dif\rho_i
&= y_i^2 - 2y_i \Exp[\rho_i] + \Exp[\rho_i^2] ,
\end{align}
where, for the expectations, $\rho_i$ has the Ricean density under the brace.
For this density, it is known that 
\begin{align}
\Exp[\rho_i]
&= \sqrt{\frac{\pi}{4\zeta^k}} \,L_{1/2}\!\left(-\zeta^k|\hat{z}_i^k|^2\right) \\
\Exp[\rho_i^2]
&= 1/\zeta^k + |\hat{z}_i^k|^2 ,
\end{align}
where the Laguerre polynomial $L_{1/2}(x)$ can be computed as
\begin{align}
L_{1/2}(x)
&= \exp\left(\frac{x}{2}\right)\left[(1-x)I_0\left(-\frac{x}{2}\right)-xI_1\left(-\frac{x}{2}\right)\right] .
\end{align}
Note that, for reasons of numerical precision, $\exp(x/2)I_d(-x/2)$ is computed using ``\texttt{besseli($d$,$-x/2$,1)}'' in Matlab, not ``\texttt{exp($x/2$).*besseli($d,-x/2$)}.''

\vspace{-8mm}
\section{Simulations}\label{sec:sims}
\vspace{-3mm}

\begin{figure}[h]
\centering
\subfigure[$\sigma_w^2=100$]{\includegraphics[width=.49\textwidth]{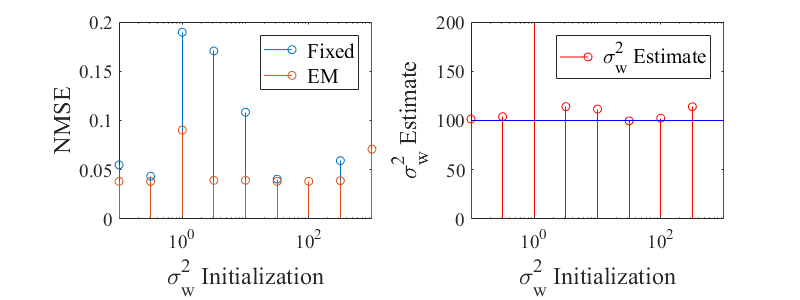}}
\subfigure[$\sigma_w^2=75$]{\includegraphics[width=.49\textwidth]{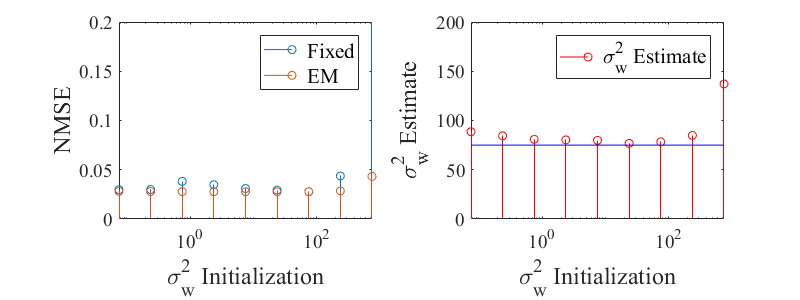}}
\subfigure[$\sigma_w^2=50$]{\includegraphics[width=.49\textwidth]{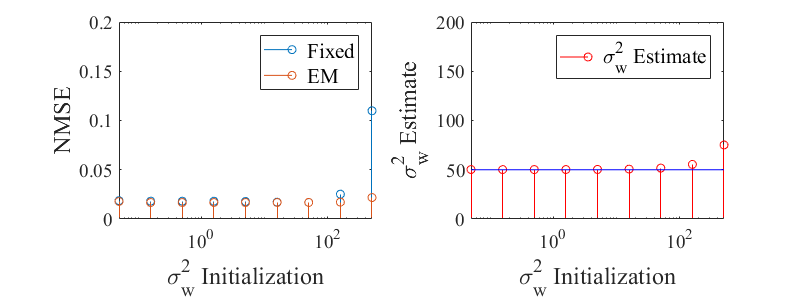}}
\subfigure[$\sigma_w^2=25$]{\includegraphics[width=.49\textwidth]{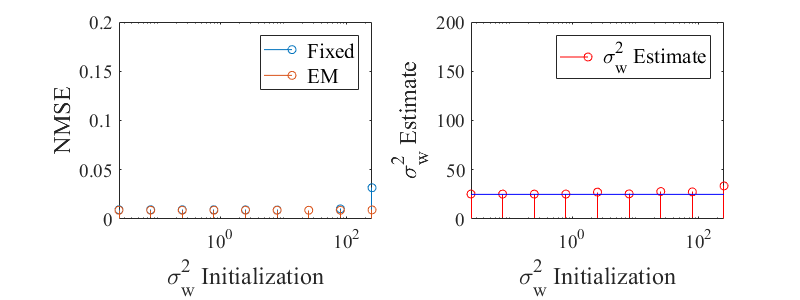}}
\caption{Reconstruction errors (left subplots) and estimates of $\sigma_w^2$ (right subplots) with different initial estimates of ${\sigma_w}^2$. The EM procedure is capable of estimating the true noise variance over a range of operating conditions. Using this estimate of the noise variance incrementally improves recovery accuracy.
}
\label{fig:ReconErrs}
\end{figure}

In this section, we demonstrate the effectiveness of the EM procedure in simulation. In particular, we show how EM can approximately recover the noise variance even when initialized by estimates far from the ground truth. This in turn enables improved signal reconstruction when the noise variance is apriori unknown.

We set up our simulations as follows. We aim to recover an i.i.d.~circular Gaussian random vector $x\in\mathbb{C}^n$, with variance $\sqrt{2}$, from phaseless noisy measurements of the form $y=|\mathbf{A}x+w|$. Our measurement matrix $\mathbf{A}$ is $8192\times 1024$ and the elements of $A$ are i.i.d.~circular Gaussian with variance $\sqrt{2}$. The elements of the noise vector $w$ also follow an i.i.d.~circular Gaussian distribution, but with variance $\sigma_w^2$. We test the cases of $\sigma_w^2=100$, $\sigma_w^2=75$, $\sigma_w^2=50$, and $\sigma_w^2=25$. prVAMP was provided with initial estimates of $\sigma_w^2$ ranging from $1\%$ to $10\times$ the true variance. Using these initializations, we reconstructed the signal with and without the EM procedure.

Figure \ref{fig:ReconErrs} presents our reconstructions. The results demonstrate that EM can be used to estimate $\sigma_w^2$. Moreover, it shows that this estimate lets prVAMP accurately reconstruct the signal even when $\sigma_w$ is not known apriori.


Code demonstrating the EM procedure is available at \url{http://gampmatlab.wikia.com/wiki/Generalized_Approximate_Message_Passing}.

\vspace{-5mm}
\section{Conclusion}
\vspace{-3mm}
This paper combines EM and GVAMP to estimate the unknown channel parameters associated with GLMs. This in turn enables GVAMP to estimate signals from their generalized linear measurements. 
In this paper we applied the proposed technique to phase retrieval and showed that it is effective at estimating unknown noise variances, thus enabling noise robust phase retrieval over a range of operating conditions.
\vspace{-3mm}

\subsection*{Acknowledgements}
Phil Schniter was supported by NSF grant CCF-1716388. 
Richard Baraniuk and Chris Metzler were supported by the DOD Vannevar Bush Faculty Fellowship N00014-18-1-2047 and the NSF GRF program, respectively. 
They were also supported by NSF grant CCF-1527501, ARO grant W911NF-15-1-0316, AFOSR grant FA9550-14-1-0088, ONR grant N00014-17-1-2551, DARPA REVEAL grant HR0011-16-C-0028, ARO grant Supp-W911NF-12-1-0407, and an ONR BRC grant for Randomized Numerical Linear Algebra.

\bibliographystyle{ieeetr}
\bibliography{./bibl.bib}
\end{document}